\documentclass[graphics,floatfix, nofootinbib,tightenlines,nobibnotes, 12pt]{revtex4-2}

\usepackage{hyperref}
\usepackage{graphicx}
\usepackage[english]{babel}
\usepackage[utf8]{inputenc}
\usepackage{amsmath,amssymb}
\usepackage{amsfonts}
\usepackage{multirow}
\usepackage{pstricks}
\usepackage{float}
\usepackage{subfigure}
\usepackage{color}
\usepackage{epsfig}
\usepackage{slashed}
\usepackage{diagbox}

\begin{document}

\title{Application of fragmentation function to the indirect production of fully charmed tetraquark}
\author{Hong-Hao Ma$^{1,2}$}
\email{mahonghao@pku.edu.cn}
\author{Zheng-Kui Tao$^{1}$}
\email{taozk@stu.gxnu.edu.cn}
\author{Juan-Juan Niu$^{1,2}$}
\email{niujj@gxnu.edu.cn, corresponding author}

\date{\today}

\begin{abstract}
The indirect production mechanisms of fully charmed tetraquark are analyzed using the NRQCD factorization and Suzuki approach, respectively. 
The process first produces a heavy charm quark through Higgs, $W^+$, or $Z^0$ decay, and then the resulting charm quark evolves into an $S$-wave fully charmed tetraquark state with quantum number $J^{PC}$, including $0^{++}$, $1^{+-}$, and $2^{++}$, via the fragmentation function. While the transverse momentum $\langle \vec{q}_T^2\rangle$ in Suzuki approach ranges from 2.01 to 299.04 GeV$^2$, the numerical results obtained from these two approaches are consistent with each other. The decay widths, branching ratios, and produced events would be predicted at LHC and CEPC, respectively. The corresponding theoretical uncertainty of heavy quark mass $m_c$ and distribution of energy fraction are also presented. The results show that the contribution for the production of $T_{4c}$ through $W^+$ decay channel at LHC is relatively large. At CEPC, a sufficient number of $T_{4c}$ events are produced through $Z^0$ decays, which is likely to be detected in future experiments.
\end{abstract}

\maketitle

\section{Introduction}

The quark model \cite{GellMann:1964nj,Zweig:1981pd,Zweig:1964jf} predicts the existence of exotic hadrons, including the tetraquarks and pentaquarks, which have more complex internal structures and are produced with a large non-perturbative effect. The research on exotics provides a good platform for studying the strong interaction and the confinement mechanism. After the first experimental observation of X(3872) as a candidate for the exotic hadron in 2003 by Belle collaboration \cite{Belle:2003nnu},
LHCb experiment in 2020 observed X(6900) from the $J/\psi$-pair production in the mass region between 6.2 and 7.4 GeV$/c^2$, covering masses of states composed of four charm quarks \cite{LHCb:2020bwg}. 
Therefore, X(6900) can be naturally regarded as a candidate for a fully charmed tetraquark. Subsequently, the experimental results of X(6900) were verified by the CMS collaboration \cite{CMS:2023owd} and ATLAS collaboration \cite{Xu:2022rnl}. 
As a special component of the tetraquark state, the fully charmed tetraquark including four heavy charm-anticharm quarks ($cc\Bar{c}\Bar{c}$), has attracted a lot of theoretical and experimental attention to deeply investigate its inner properties.
In theory, many phenomenological extensions of the quark model also predict the existence of fully charmed tetraquarks, i.e. $T_{cc\Bar{c}\Bar{c}}$, labeled $T_{4c}$ for convenience. In order to better understand the nature of these fully charmed tetraquarks, research on the production, decay and mass spectra of $T_{4c}$ has been investigated \cite{Feng:2023agq,Feng:2023ghc,Sang:2023ncm,Becchi:2020uvq,Bedolla:2019zwg,Zhao:2020jvl,Chen:2020xwe}. Theoretically, the structure of such a tetraquark is usually assumed to be a $\langle cc \rangle$ diquark and a $\Bar{c}\Bar{c}$ antidiquark attracted to each other by color interactions. The $\langle cc \rangle$ diquark in color antitriplet acting as an antiquark has a diquark-antiquark symmetry (DAS) \cite{Cheng:2020wxa}. Therefore $T_{4c}$ can be considered as a color-singlet state formed by the compact diquark-antidiquark. Other alternative dynamical mechanisms have also been proposed, such as a meson molecule \cite{Tornqvist:1993ng,Guo:2013sya,Wang:2013daa,Guo:2017jvc} or a hadroquarkonium \cite{Dubynskiy:2008mq,Ferretti:2020ewe}.

The fragmentation picture, based on collinear factorization, is also applied to study the production mechanism of fully charmed tetraquark at LHC, especially gluon-to-$T_{4c}$ fragmentation function \cite{Feng:2020riv,Celiberto:2024mab} and charm-to-$T_{4c}$ fragmentation function \cite{Celiberto:2024mab,Bai:2024ezn}
within nonrelativistic QCD (NRQCD) framework \cite{Bodwin:1994jh,Petrelli:1997ge}. The fragmentation function of a given parton (charm or gluon) to a heavy flavored $T_{4c}$ can be factorized into the convolution of the short-distance coefficients and the long-distance matrix elements with an initial factorization energy-scale \cite{Mele:1990yq,Mele:1990cw,Cacciari:1993mq,Cacciari:1996wr,Kniehl:2005mk}, in which the short-distance coefficients are perturbatively calculable based on the high-energy hard process. On the other hand, the long-distance matrix elements are non-perturbative and can be related to the $T_{4c}$ four-body wave functions at the origin, which have been evaluated by phenomenological potential models. Another fragmentation picture, based on collinear factorization, has also been proposed for charm-to-$T_{4c}$, called Suzuki approach \cite{Suzuki:1985up}. It builds on spin-physics, Ref. \cite{Nejad:2021mmp} introduces the calculation of the Suzuki-driven, initial-scale input of the [$Q,\bar{Q}\to X_{Qq\bar{Q}\bar{q}}$] fragmentation function, whereas its threshold-consistent Dokshitzer-Gribov-Lipatov
Altarelli-Parisi (DGLAP) \cite{Dokshitzer:1977sg,Altarelli:1977zs,Curci:1980uw,Furmanski:1980cm} evolution and first application of phenomenology to the production of heavy-light tetraquarks ($Qq\bar{Q}\bar{q}$) was in Refs. \cite{Celiberto:2023rzw,Celiberto:2024mrq}, and then it was well-adapted to the $T_{4c}$ case in Ref. \cite{Celiberto:2024mab}. The fragmentation function in Suzuki approach contains a model parameter, the average transverse momentum $\langle \vec{q}_T^2\rangle$ of the hadron relative to the jet, to describe its fragmentation effect.

The hadronic production of fully charmed tetraquarks at LHC \cite{Feng:2023agq} and photoproduction at electron-ion
colliders \cite{Feng:2023ghc} have been investigated separately. The results show that the cross section and
the production rate are sizable. In this paper, $S$-wave fully charmed tetraquark $T_{4c}$ will be studied using charm-to-$T_{4c}$ fragmentation function at LHC and CEPC, respectively. At leading order, the indirect production mechanism is that Higgs and weak bosons first decay into a quark-antiquark pair ($c\bar{c}$ or $c\bar{s}$) and then the resulting charm quark hadronizes into an $S$-wave fully charmed tetraquark state through the fragmentation function. The $J^{PC}$ quantum numbers of fully charmed tetraquark state contain $0^{++}$, $1^{+-}$, and $2^{++}$.

The paper is organized as follows. In Sec. II, we present the calculation strategy of NRQCD factorization and Suzuki approach. Sec. III is devoted to numerical results of decay widths, branching ratios, phenomenological predictions about the produced events of $T_{4c}$ at LHC and CEPC, the differential distribution over the energy fraction $z$, and the corresponding theoretical uncertainty. Finally, a summary is given in Sec. IV.

\section{Calculation Technology}

Using fragmentation function approach, the decay width of the indirect processes through Higgs, $W^{+}$, and $Z^{0}$ decay for the production of $T_{4c}$ at leading order can be factorized as
\begin{eqnarray}\label{}
\Gamma(Higgs /W^{+}/Z^{0}(p_0)&& \rightarrow T_{4c}(p_1)+ \bar {c}/\bar{s}(p_2) )  \nonumber\\
&& =\hat{\Gamma}(Higgs /W^{+}/Z^{0} \rightarrow  c +\bar {c}/\bar{s} ) 
 \int_{0}^{1}d{z}D _{c \to T_{4c}}(z,\mu),
\end{eqnarray}
in which the energy fraction $z=\frac{E_{T_{4c}}}{E_{T_{4c}}^{max}}$, the factorization scale $\mu=5m_c$, $D _{c \to T_{4c}}(z,\mu)$ is the fragmentation function for the production of $T_{4c}$, which refers to the probability for a charm quark at the initial-scale $\mu$ to fragment into a $T_{4c}$ carrying away a energy fraction $z$ of its momentum. Associated with the production of the charm quark, there are two kinds of decay processes, $Higgs/Z^{0} \rightarrow  c +\bar {c}$ and $W^{+} \rightarrow  c +\bar{s}$ in the decay width $\hat{\Gamma}(Higgs/W^{+}/Z^{0} \rightarrow  c +\bar {c}/\bar{s} )$.
It should be noted that the decay channel $W^{+} \rightarrow  c +\bar{b}$ is negligible due to the suppression of CKM matrix element $|V_{cb}|$ $(|\frac{V_{cb}}{V_{cs}}|^2 < 0.002)$.

\subsection{NRQCD Factorization}

According to the factorization theorem, the fragmentation function $D _{ c \to T_{4c}}(z,\mu)$ is independent of the hard processes in which the heavy charm quark is produced.
And the charm-to-$T_{4c}$ fragmentation function can
be expressed as the convolution of the short-distance coefficients and the long-distance matrix elements, that is 

\begin{eqnarray}\label{frag}
D _{ c \to T_{4c}}(z) 
=\sum_n d_n(z)\langle O_n^{J^{PC}}\rangle,
\end{eqnarray}
where $d_n(z)$ stands for the perturbative short-distance coefficients with color quantum number $n$, $\langle O_n^{J^{PC}}\rangle$ is the long-distance matrix elements to hadronization with quantum number $J^{PC}$, which can be $0^{++}$, $1^{+-}$ and $2^{++}$ for $S$-wave tetraquarks. For the internal color interaction of the fully charmed tetraquark, the color-singlet tetraquark can be decomposed by the color $\langle cc\rangle$-$\langle \bar{c}\bar{c}\rangle$ diquark-antidiquark basis. 
The color configuration of $\langle cc\rangle$ diquark can be $\bf 3\bigotimes 3 = \bar{3}\bigoplus{6}$ for the decomposition of SU$(3)_c$ color group. Due to the Fermi-Dirac statistics, the diquark in color antitriplet corresponds to spin 1, while that in color sextuplet corresponds to the spin 0. In this way, the color-singlet $T_{4c}$ can be decomposed as $\bf \bar{3}\bigotimes 3$ or $\bf 6 \bigotimes \bar{6}$ configurations. The 
$\bf \bar{3}\bigotimes 3$ case can have spin 0, 1 and 2, while the $\bf 6 \bigotimes \bar{6}$ one can only have spin 0. Therefore, the fragmentation function in Eq.~(\ref{frag}) at lowest
 order in velocity can be expanded as
\begin{eqnarray}\label{}
D _{ c \to T_{4c}}(z) 
=\frac{d_{3,3}[c\to cc\bar{c}\bar{c}^{J^{PC}}]}{m_c^9} \langle O_{3,3}^{J^{PC}}\rangle + \frac{d_{6,6}[c\to cc\bar{c}\bar{c}^{J^{PC}}]}{m_c^9} \langle O_{6,6}^{J^{PC}}\rangle \nonumber \\ 
+ \frac{d_{3,6}[c\to cc\bar{c}\bar{c}^{J^{PC}}]}{m_c^9} 2\rm {Re}\langle \it O_{\rm 3,6}^{J^{PC}}\rangle + \dots.
\end{eqnarray}

We shall directly use the short-distance coefficients $d_n(z)$ obtained by the decay of the charm quark to perform the numerical calculation, the results of which are given in equations (5.1-5.5) of Ref. \cite{Bai:2024ezn}. As for the long-distance matrix element, it can be related to the wave functions at the origin, which can be estimated by Lattice QCD \cite{Lepage:1992tx,Hughes:2017xie}, QCD sum rules \cite{Yang:2020wkh,Wang:2014gwa} or phenomenological potential models \cite{ Eichten:1994gt,Debastiani:2017msn,Giron:2020wpx,Jin:2020jfc,Gordillo:2020sgc}. However, the results of Lattice and sum rules on the fully charmed tetraquark are very few.
The wave functions at the origin obtained by five distinct potential models can be used to calculate the long-distance matrix element for the production of $T_{4c}$, including its radially excited states up to the 2$S$ level \cite{Lu:2020cns,Zhao:2020nwy,liu:2020eha,Yu:2022lak,Wang:2019rdo}. The results are listed in Table~\ref{wf} and show that the four-body Schr$\mathrm{\ddot o}$dinger wave functions at the origin predicted by five different potential models are significantly different. The theoretical uncertainty caused by the long-distance matrix elements in the non-perturbative region is greater.

\begin{table}[htb]
\centering
\begin{tabular}{|c|c|ccccc|}
\hline
\multirow{2}{*}{n$S$}&
 \multirow{2}{*}{$\langle {O_n^{{J^{PC}}}}\rangle $} &
  \multicolumn{5}{c|}{Models} \\ \cline{3-7} 
\multicolumn{1}{|c|}{} &
   &
  \multicolumn{1}{c|}{I\cite{Lu:2020cns}} &
  \multicolumn{1}{c|}{II\cite{Zhao:2020nwy}} &
  \multicolumn{1}{c|}{III\cite{liu:2020eha}} &
  \multicolumn{1}{c|}{IV\cite{Yu:2022lak}} &
 V\cite{Wang:2019rdo} \\ \hline
\multirow{5}{*}{1$S$}  &$\langle {O_{6,6}^{ 0 ^{ +  + }}} \rangle $
   &
  \multicolumn{1}{c|}{0.0128} &
  \multicolumn{1}{c|}{2.50} &
  \multicolumn{1}{c|}{0.0027} &
  \multicolumn{1}{c|}{0.0173} &
  0.00226 \\ \cline{2-7} 
  &$\langle {O_{3,6}^{ 0^{ +  + }}} \rangle $
   &
  \multicolumn{1}{c|}{0.0211} &
  \multicolumn{1}{c|}{3.65} &
  \multicolumn{1}{c|}{0.0033} &
  \multicolumn{1}{c|}{-0.0454} &
  0.00215 \\ \cline{2-7} 
  &$\langle {O_{3,3}^{ 0^{ +  + }}} \rangle $
   &
  \multicolumn{1}{c|}{0.0347} &
  \multicolumn{1}{c|}{5.33} &
  \multicolumn{1}{c|}{0.0041} &
  \multicolumn{1}{c|}{0.119} &
  0.00204 \\ \cline{2-7} 
 &$\langle {O_{3,3}^{1^{ +  - }}} \rangle $
   &
  \multicolumn{1}{c|}{0.0780} &
  \multicolumn{1}{c|}{12.6} &
  \multicolumn{1}{c|}{0.011} &
  \multicolumn{1}{c|}{0.0975} &
  0.00876 \\ \cline{2-7} 
 &$\langle {O_{3,3}^{2^{ +  + }}} \rangle $
   &
  \multicolumn{1}{c|}{0.072} &
  \multicolumn{1}{c|}{13.6} &
  \multicolumn{1}{c|}{0.012} &
  \multicolumn{1}{c|}{0.254} &
  0.0117 \\ \hline\hline
\multirow{5}{*}{2$S$} 
&$\langle {O_{6,6}^{0^{ +  + }}} \rangle $
   &
  \multicolumn{1}{c|}{0.0347} &
  \multicolumn{1}{c|}{5.46} &
  \multicolumn{1}{c|}{0.0058} &
  \multicolumn{1}{c|}{0.0179} &
  0.000545 \\ \cline{2-7} 
   &$\langle {O_{3,6}^{0^{ +  + }}} \rangle $
   &
  \multicolumn{1}{c|}{0.0538} &
  \multicolumn{1}{c|}{10.6} &
  \multicolumn{1}{c|}{0.0067} &
  \multicolumn{1}{c|}{0.0386} &
  0.000890 \\ \cline{2-7} 
   &$\langle {O_{3,3}^{0^{ +  + }}} \rangle $
   &
  \multicolumn{1}{c|}{0.0832} &
  \multicolumn{1}{c|}{20.5} &
  \multicolumn{1}{c|}{0.0077} &
  \multicolumn{1}{c|}{0.0832} &
  0.00145 \\ \cline{2-7} 
&$\langle {O_{3,3}^{1^{ +  + }}} \rangle $
   &
  \multicolumn{1}{c|}{0.1887} &
  \multicolumn{1}{c|}{32.5} &
  \multicolumn{1}{c|}{0.021} &
  \multicolumn{1}{c|}{0.0648} &
  0.0173 \\ \cline{2-7} 
 &$\langle {O_{3,3}^{2^{ +  + }}} \rangle $
   &
  \multicolumn{1}{c|}{0.1775} &
  \multicolumn{1}{c|}{30.5} &
  \multicolumn{1}{c|}{0.026} &
  \multicolumn{1}{c|}{0.151} &
  0.0236 \\ \hline
\end{tabular}
\caption{The long-distance matrix elements $\langle O_{n}^{J^{PC}}\rangle$ for the production of $T_{4c}$ in n$S$-wave using the wave functions at the origin obtained by five distinct potential models.}
\label{wf}
\end{table}

In higher-order calculations, there often appear final-state collinear singularities, which, however, can be reabsorbed in the definition of the fragmentation function. This generates a dependence of fragmentation function on the factorization energy-scale, regulated by the DGLAP
evolution. After applying DGLAP to evolve the energy-scale $\mu$ into the considered decaying boson, $m_H$, $m_W^{+}$ or $m_Z$ correspond to three different decay processes, collinear singularities are no more, but large energy
logarithms, such as In$(M_{T_{4c}}/E_{T_{4c}})$, are present and can be resummed to all orders,
\begin{eqnarray}\label{}
\mu \frac{\partial}{\partial\mu}  D _{ c \to T_{4c}}(z,\mu) 
=\frac{\alpha_s(\mu)}{\pi}\sum_{i\in\{c,g\}}\int_{z}^{1}\frac{dy}{y}P_{c\to i}(z/y)D _{ i \to T_{4c}}(y,\mu),
\end{eqnarray}
in which the quark-to-quark and quark-to-gluon splitting kernels $P_{q\to q}(z)$ and $P_{q\to g}(z)$ at leading order in the QCD running coupling, $\alpha_s$, are
\begin{eqnarray}\label{}
P_{q\to q}(z)&=&C_F  \left [\frac{1+z^2}{(1-z)_+}+\frac{3}{2}\delta(1-z) \right],\nonumber\\
P_{q\to g}(z)&=&C_F  \frac{1+(1-z)^2}{z}.
\end{eqnarray}
The exact solution of the DGLAP equations can be performed only numerically via codes like APFEL(++) \cite{Bertone:2013vaa}, EKO \cite{Candido:2022tld} or FFEVOL \cite{Hirai:2011si}. An approximate analytical method can also be available to solve the DGLAP equation \cite{Field:1989uq,Zheng:2015ixa}. Thus, the fragmentation functions can be evolved to the energy-scale of $m_H$, $m_W^{+}$, or $m_Z$, corresponding to the three decay processes.

\begin{figure}
    \centering
    \includegraphics[width=0.5\linewidth]{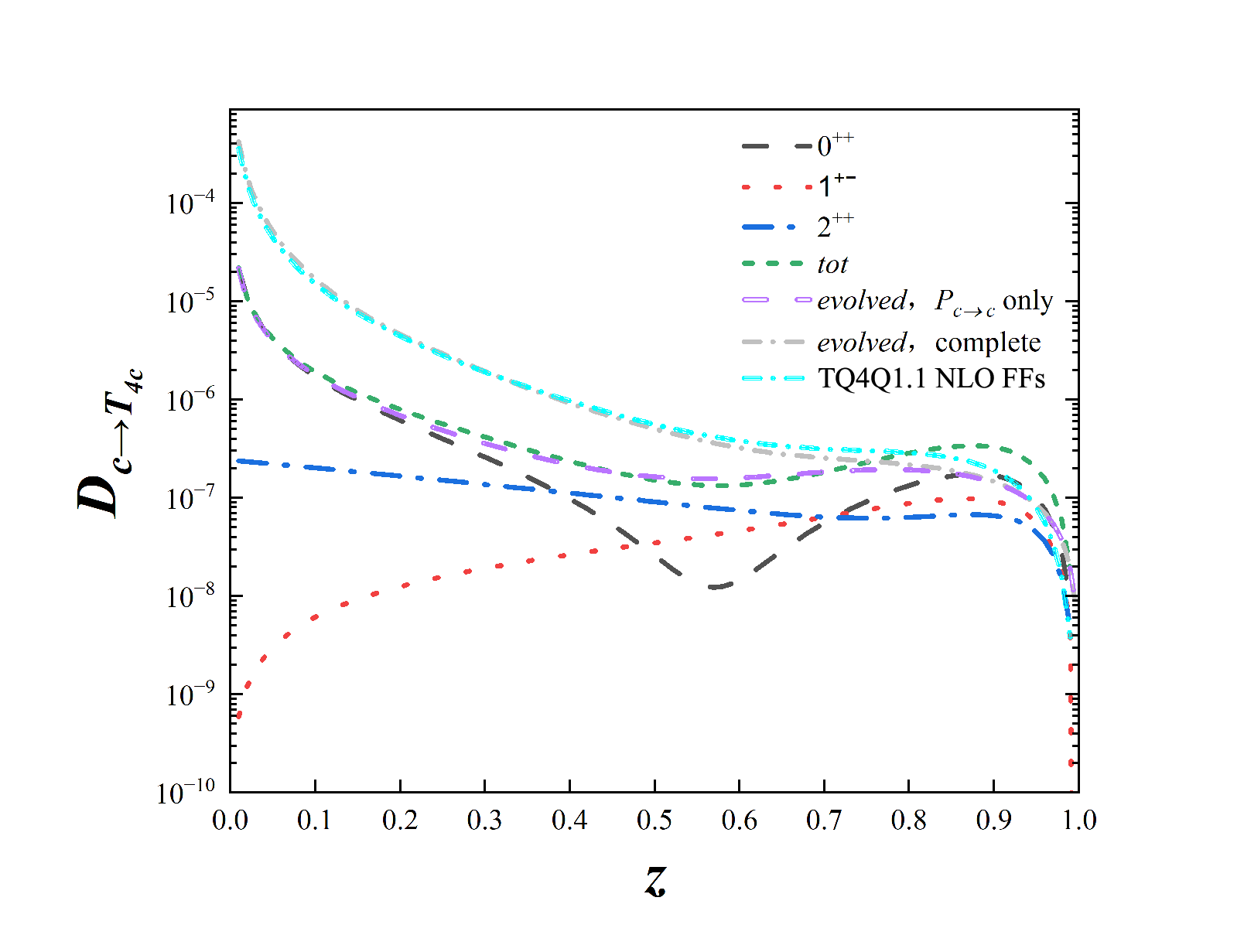}
    \hspace{-.60in}
    \includegraphics[width=0.5\linewidth]{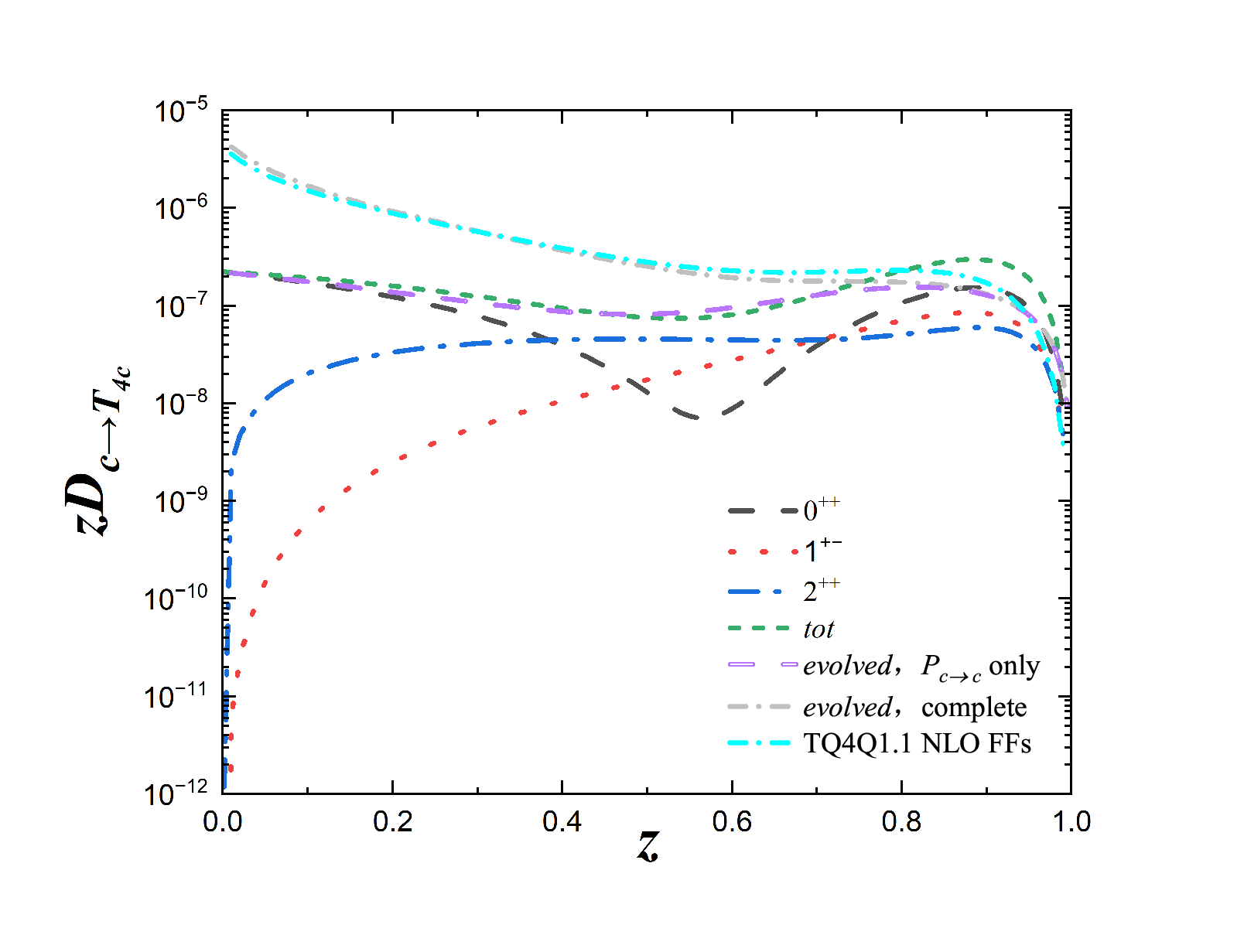}
    \caption{Energy fraction distributions of $D_{ c \to T_{4c}}(z)$ and $zD_{ c \to T_{4c}}(z)$ with different quantum number $J^{PC}$, the total and the evolved fragmentation functions.}
    \label{fff}
\end{figure}

To analyze the trend of behavior, the energy fraction distributions of the fragmentation function $D_{ c \to T_{4c}}(z)$ and $zD_{ c \to T_{4c}}(z)$ with different quantum numbers $J^{PC} = 0^{++},~1^{+-},~2^{++}$ are plotted separately in Fig.~\ref{fff}. Each of them has its own unique behavior trend. After summing the different quantum numbers $J^{PC} = \rm{tot}~(0^{++}+1^{+-}+2^{++})$, we get the distribution of the total fragmentation function $D_{ c \to T_{4c}}(z)$ ($tot$), which is also shown in Fig.~\ref{fff}. After energy-scale evolution using DGLAP equation, the evolved one containing only the contribution of $P_{q\to q}(z)$ is slightly different from $tot$, except for the large $z$ region. In order to give a more complete theoretical prediction, with the help of FFEVOL for the resummation of the DGLAP equation at NLO, the energy-scale evolution including $P_{q\to q}(z)$ and $P_{q\to g}(z)$ is added in Fig.~\ref{fff}, which is different from the distribution of that only involving $P_{c\to c}$, especially in the small $z$ region. Here the initial-scale 5$m_c$ are evolved to $m_Z$, and the energy-scale evolutions to $m_H$ and $m_W^{+}$ are very close to those of $m_Z$ and will not be shown.

\subsection{Suzuki Approach}

Another approach for the fragmentation function of charm-to-$T_{4c}$ in color-singlet $S$-wave state is calculated in Suzuki model \cite{Celiberto:2024mab}, which is spin-dependent. Ref. \cite{Nejad:2021mmp} introduces the calculation of the Suzuki-driven, initial-scale input of the [$Q,\bar{Q}\to X_{Qq\bar{Q}\bar{q}}$] FF, whereas its threshold-consistent DGLAP evolution. Suzuki approach is an analogous factorization scheme to some extent as the NRQCD factorization. The production mechanism in the Suzuki approach is that a four-quark $(cc\bar{c}\bar{c})$ system is first perturbatively produced by an outgoing charm quark emission, and then non-perturbatively hadronic related to the wave function. The fragmentation function $D _{ c \to T_{4c}}$ in Suzuki approach relies on a model parameter, the transverse momentum $\langle \vec{q}_T^2\rangle$ of the hadron relative to the jet, to describe its fragmentation effect \cite{Suzuki:1985up}, the expression can be written as

\begin{eqnarray}\label{}
D _{ c \to T_{4c}}(z) 
=\mathcal{N}_{c}\frac{(1-z)^5}{\Delta^2(z)} \sum _{k=0}^{3} \gamma_k^{(c)}(z) z^{2(k+2)} \Big\{\frac{\langle \vec{q}_T^2\rangle}{m_c^2} \Big \}^k,
\end{eqnarray}
where $\mathcal{N}_{c}=[512\pi^2 C_F f_\mathcal{B} \alpha_s(2m_c)^2]^2$ with $C_F  \equiv \frac{N_c^2 -1}{2N_c}$, the hadron decay constant $f_\mathcal{B}=0.25$ GeV \cite{ParticleDataGroup:2020ssz}, and

\begin{eqnarray}
\frac{1}{\Delta(z)}&=&\frac{m_c^8}{[(4-3z)^2m_c^2+z^2 \langle\vec{q}_T^2\rangle]^2[(4-z)^2m_c^2+z^2\langle\vec{q}_T^2\rangle][(1-z)M_{T_{4c}}^2
 +z^2(m_c^2+\langle\vec{q}_T^2\rangle)]}            ,\nonumber\\
\gamma_0^{(c)}(z)&=&(4-z)^2(256-512z+416z^2-160z^3+33z^4),\nonumber\\
\gamma_1^{(c)}(z)&=& 768-1280z+1248z^2-464z^3+67z^4,\nonumber\\
\gamma_2^{(c)}(z)&=& 48-40z+35z^2,\nonumber\\
\gamma_3^{(c)}(z)&=&1.
\end{eqnarray}

For a numerical phenomenological analysis of the fragmentation function $D _{c \to T_{4c}}$ in the Suzuki approach, the transverse momentum $\langle \vec{q}_T^2\rangle$ can simply be replaced by its average value. Based on the Suzuki approach, the first family of DGLAP-evolving, collinear fragmentation function for $T_{4c}$ tetraquarks, named TQ4Q1.0, was derived in Ref. \cite{Celiberto:2024mab}. It embodies not only the initial-scale input for the charm quark taken from Suzuki \cite{Suzuki:1985up,Nejad:2021mmp}, but also the gluon one \cite{Feng:2020riv} inspired by potential NRQCD. The inclusion of the NRQCD input for the heavy-quark channel \cite{Bai:2024ezn} and the extension to the $T_{4b}$ case were done in Ref. \cite{Celiberto:2024beg}, by deriving the TQ4Q1.1 fragmentation function family. 

For comparison, the results obtained by TQ4Q1.1 (TQ4Q1.1 NLO FFs) are also presented in Fig.~\ref{fff}.
Similar to the results of the complete evolution of NRQCD (evolved, complete), the only difference lies in the selection of initial energy-scale. In TQ4Q1.1, the initial energy-scale of charm-to-$T_{4c}$ is 5$m_c$ and is 4$m_c$ for gluon-to-$T_{4c}$, while the initial energy-scale of complete evolution (evolved, complete) is 5$m_c$.

\begin{figure}
    \centering
    \includegraphics[width=0.5\linewidth]{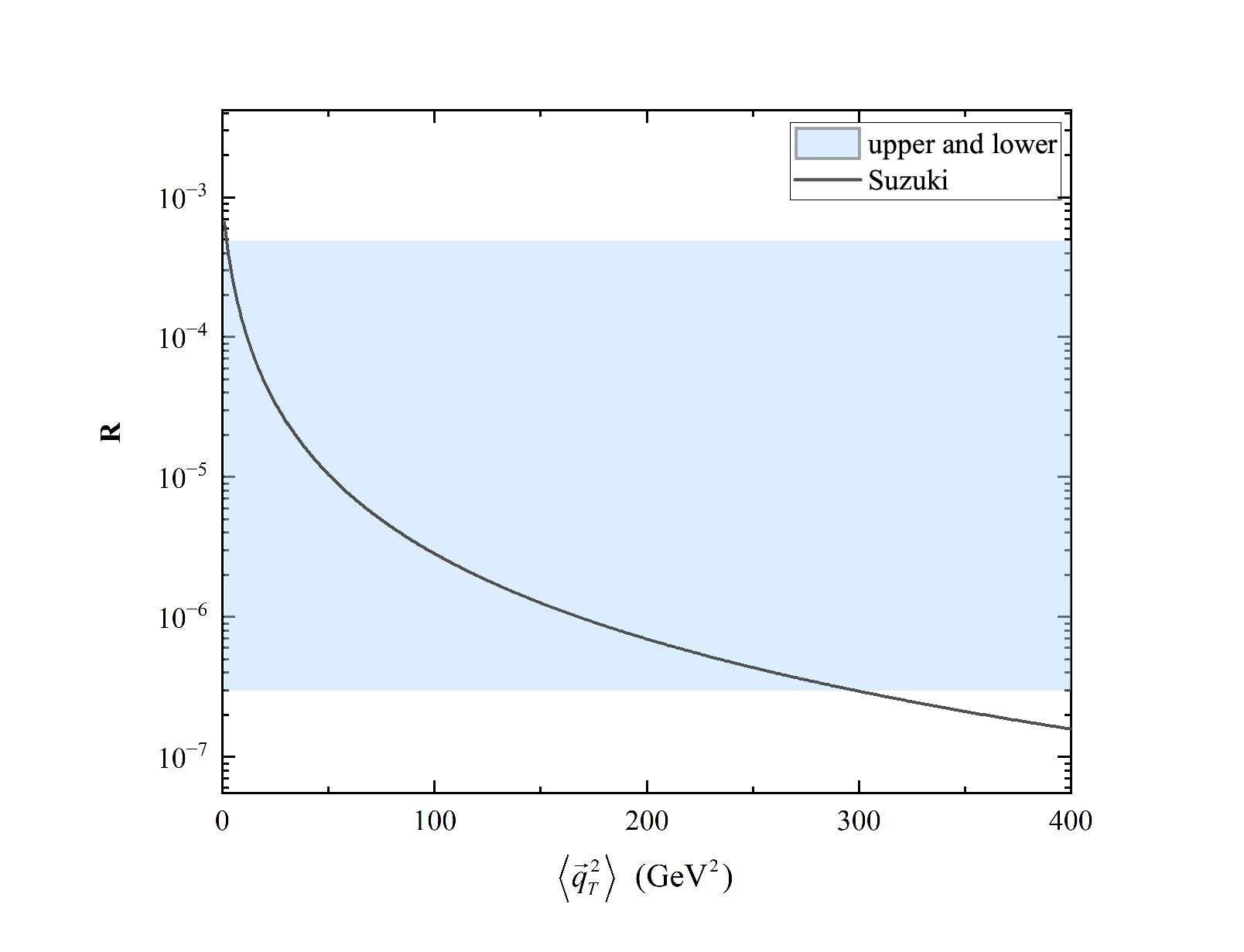}
    \caption{Transverse momentum $\langle \vec{q}_T^2\rangle$ distribution of $R$ in Suzuki approach with the results $R$ obtained in NRQCD factorization.}
    \label{tmd}
\end{figure}

Then the transition probability $R$ of the fragmentation function in Suzuki approach can be obtained by integrating the energy fraction $z$. The fragmentation function in Suzuki approach does not distinguish states with different quantum number $J^{PC}$, so we conduct a simple comparative analysis with the results of $J^{PC}=tot~(0^{++}+ 1^{+-}+2^{++})$ in NRQCD factorization. The transverse momentum $\langle \vec{q}_T^2\rangle$ distribution of $R$ in Suzuki approach is shown in Fig.~\ref{tmd}, which also includes the $R$ obtained by NRQCD factorization. The lower limit in Fig.~\ref{tmd} is the value of $R$ in NRQCD factorization with Model V, and the upper limit is the result of model II. From Fig.~\ref{tmd}, one can see that when the transverse momentum $\langle \vec{q}_T^2\rangle$ range in Suzuki approaches is 2.01 to 299.04 GeV$^2$, the numerical results of these two approaches are consistent with each other.

\section{Numerical Results}

The input parameters used in the numerical calculation for the indirect production of $T_{4c}$ at LHC and CEPC are listed as follows

\begin{eqnarray}
&&m_c=1.5~\rm{GeV},~~~~~~~~~~~~~\it{m_s}=\rm 0~{GeV},~~~~~~~~~~\rm{cos} \theta_W=\it m_W / m_Z,\nonumber\\
&&\it{m_W}=\rm 80.385~{GeV},~~~~~~~\it{m_Z}=\rm 91.1876~\rm{GeV},~\it{m_H}=\rm 125.18~{GeV}, \nonumber\\
&&\Gamma_{\it H}=4.2~\rm{MeV},~~~~~~~~~~~~~\Gamma_{\it W^+}=~{2.085~\rm GeV},~\Gamma_{\it Z^0}=2.4952~{\rm GeV},\nonumber\\
&&G_{F}=1.1663787 \times 10^{-5},~~\alpha_{s}(\it{m_Z})=\rm 0.118,~~~~~\alpha_{s}( 5\it{m_c})=\rm 0.191,
\label{parameters}
\end{eqnarray}
where the quark masses are the same as Ref.~\cite{Baranov:1995rc} and the others can be obtained from PDG~\cite{ParticleDataGroup:2020ssz}. Programs FeynArts 3.9 \cite{Hahn:2000kx}, FeynCalc 9.3 \cite{Shtabovenko:2020gxv} and the modified FormCalc \cite{Hahn:1998yk} were used to perform the algebraic and numerical calculations.

To estimate the produced events of $T_{4c}$ using formula $N_{tot} \rm {Br}_{Higgs/\it W^+/Z^0 \to T_{\rm 4 \it c}}$, where Br$_{Higgs/W^+/Z^0 \to T_{4c}}$ is the corresponding branching ratio of $\frac{\Gamma_{H/W^+/Z^0 \to T_{4c}}}{\Gamma_{H/W^+/Z^0}}$, the total events $N_{tot}$ of Higgs, $W^+$ and $Z^0$ produced at LHC and CEPC are needed~\cite{Niu:2024ghc}. At LHC with luminosity $\mathcal{L} \propto10^{34}~\rm cm^{-2} s^{-1}$ and collision energy $\sqrt{s}$ = 14 TeV, there could be $1.65\times 10^8$ Higgs events and $3.07\times10^{10}$ $W^+$, and $1.00\times 10^9$ $Z^0$ boson produced per year \cite{Higgsevents,Gaunt:2010pi,Qiao:2011yk,Liao:2015vqa};
CEPC running at $\sqrt{s}$ = 240 GeV corresponding to an integrated $\mathcal{L}=2.2~\rm ab^{-1}$ could produce $4.30\times 10^5$ of Higgs events per year; When $\sqrt{s}$ is 160 GeV and integrated $\mathcal{L}= 6.9~\rm ab^{-1}$ at CEPC, there would be $2.10\times 10^8$ $W^+$ events produced per year; As for the produced events of $Z^0$ at CEPC per year, there would be $2.05\times 10^{12}$ when it would be operated as a Super-Z factor at $\sqrt{s}=91$ GeV and integrated $\mathcal{L}= 50~\rm ab^{-1}$ \cite{CEPCStudyGroup:2023quu,Ai:2024nmn}. 

\subsection{Main results}\label{basic}

The numerical results of the fragmentation probability $R=\int_{0}^{1}d{z} D _{c \to T_{4c}^{J^{PC}}}(z)$ with different quantum numbers $J^{PC} = 0^{++}, 1^{+-}, 2^{++}$, and tot $(0^{++}+1^{+-}+2^{++})$, are presented, respectively, in Table~\ref{fp} corresponding to five potential models mentioned above. Based on this situation, the decay widths of three processes Higgs$/W^ +/Z^0 \to {T_{4c}} + X$ in five distinct models with $J^{PC} = 0^{++},~1^{+-},~2^{++}$, and tot $(0^{++} +1^{+-} +2^{++})$, can be estimated respectively and are presented in Table~\ref{tdw}. The corresponding branching ratios and produced events of $T_{4c}$ through these three decay channels can also be predicted in Tables~\ref{wdw}-\ref{zdw}. To deal with Br, the partial decay widths for Higgs, $W^+$, and $Z^{0}$ decay into a pair of quark-antiquark are calculated, and the results are $\hat{\Gamma}(Higgs\rightarrow c +\bar {c})=1.8443 \times 10^{ - 4}$~GeV, $\hat{\Gamma}(W^{+}\rightarrow c +\bar {s})=6.8146 \times 10^{ - 1}$~GeV, and $\hat{\Gamma}(Z^{0}\rightarrow c +\bar {c})=2.8936 \times 10^{- 1}$~GeV. However, the decay widths and predicted produced events of $T_{4c}$ by Higgs decay are very small, even negligible, to be discussed compared to that through $W^+$ and $Z^0$ decay. 

Tables~\ref{tdw}-\ref{zdw} show that the results for the production of 1$S$-wave $T_{4c}$ in five different models can differ by 3-4 orders of magnitude, with the largest result coming from model II and the smallest from model V.
 The largest contribution of branching ratios for $T_{4c}$ production is via $W^+$ decay, which is $9.74\times 10^{-8}\sim 1.61\times 10^{-4}$. Given that the produced number of $T_{4c}$ events at LHC through $W^+$ decay is $3.00\times10^3 \sim 4.93\times 10^6$, there is some potential to observe it at LHC. Due to the abundance of $Z^0$-events produced at CEPC, the predicted number of $T_{4c}$ events through $Z^0$ decay is very promising, can be $10^{4-8}$ with different potential models.

\begin{table}[htb]
\centering
\begin{tabular}{|cccccc|}
\hline
\multicolumn{1}{|c|}{Models} & \multicolumn{1}{c|}{I} & \multicolumn{1}{c|}{II} & \multicolumn{1}{c|}{III} & \multicolumn{1}{c|}{IV} & V \\ \hline
\multicolumn{6}{|c|}{1S}                                                                                                   \\ \hline
\multicolumn{1}{|c|}{${R^{0^{ +  + }}}$} & \multicolumn{1}{c|}{$2.72 \times {10^{ - 6}}$} & \multicolumn{1}{c|}{$4.64 \times {10^{ - 4}}$} & \multicolumn{1}{c|}{$4.18 \times {10^{ - 7}}$ } & \multicolumn{1}{c|}{$8.68 \times {10^{ - 7}}$ } & $2.76 \times {10^{ - 7}}$  \\ \hline
\multicolumn{1}{|c|}{${R^{1^{ +  - }}}$} & \multicolumn{1}{c|}{$3.98 \times {10^{ - 8}}$} & \multicolumn{1}{c|}{$6.42 \times {10^{ - 6}}$} & \multicolumn{1}{c|}{$5.61 \times {10^{ - 9}}$} & \multicolumn{1}{c|}{$4.97 \times {10^{ - 8}}$ } & $4.46 \times {10^{ - 9}}$ \\ \hline
\multicolumn{1}{|c|}{${R^{2^{ +  + }}}$} & \multicolumn{1}{c|}{$1.10 \times {10^{ - 7}}$} & \multicolumn{1}{c|}{$2.08 \times {10^{ - 5}}$ } & \multicolumn{1}{c|}{$1.84 \times {10^{ - 8}}$} & \multicolumn{1}{c|}{$3.89 \times {10^{ - 7}}$} & $1.79 \times {10^{ - 8}}$  \\ \hline
\multicolumn{1}{|c|}{${R^{\text{tot}}}$} & \multicolumn{1}{c|}{$2.87 \times {10^{ - 6}}$} & \multicolumn{1}{c|}{$4.92 \times {10^{ - 4}}$} & \multicolumn{1}{c|}{$4.42 \times {10^{ - 7}}$ } & \multicolumn{1}{c|}{$1.31 \times {10^{ - 6}}$} & $2.98 \times {10^{ - 7}}$ \\ \hline \hline
\multicolumn{6}{|c|}{2S}                                                                         \\ \hline
\multicolumn{1}{|c|}{${R^{0^{ +  + }}}$}  & \multicolumn{1}{c|}{$6.88 \times {10^{ - 6}}$} & \multicolumn{1}{c|}{$1.41 \times {10^{ - 3}}$ } & \multicolumn{1}{c|}{$8.46 \times {10^{ - 7}}$} & \multicolumn{1}{c|}{$5.27 \times {10^{ - 6}}$ } & $1.15 \times {10^{ - 7}}$  \\ \hline
\multicolumn{1}{|c|}{${R^{1^{ +  - }}}$}  & \multicolumn{1}{c|}{ $9.62 \times {10^{ - 8}}$} & \multicolumn{1}{c|}{$1.66 \times {10^{ - 5}}$ } & \multicolumn{1}{c|}{$1.07 \times {10^{ - 8}}$ } & \multicolumn{1}{c|}{$3.30 \times {10^{ - 8}}$} & $8.82 \times {10^{ - 9}}$ \\ \hline
\multicolumn{1}{|c|}{${R^{2^{ +  + }}}$}  & \multicolumn{1}{c|}{$2.72 \times {10^{ - 7}}$} & \multicolumn{1}{c|}{$4.67 \times {10^{ - 5}}$ } & \multicolumn{1}{c|}{$3.98 \times {10^{ - 8}}$ } & \multicolumn{1}{c|}{$2.31 \times {10^{ - 7}}$} &  $3.61 \times {10^{ - 8}}$ \\ \hline
\multicolumn{1}{|c|}{${R^{\text{tot}}}$}  & \multicolumn{1}{c|}{$7.25 \times {10^{ - 6}}$} & \multicolumn{1}{c|}{$1.47 \times {10^{ - 3}}$} & \multicolumn{1}{c|}{ $8.96 \times {10^{ - 7}}$} & \multicolumn{1}{c|}{$5.54 \times {10^{ - 6}}$ } & $1.60 \times {10^{ - 7}}$  \\ \hline
\end{tabular}
\caption{The fragmentation probabilities $R$ with $J^{PC} = 0^{ +  + }, 1^{ +  - }, 2^{ +  + }$, and tot $(0^{ +  + } + 1^{ +  - } + 2^{ +  + })$ corresponding to five potential models.}
\label{fp}
\end{table}

\begin{table}[htb]
\centering
\begin{tabular}{|c|c|c|c|c|c|c|}
\hline
\multicolumn{2}{|c|}{\diagbox{$J^{PC}$}{$\Gamma$ (GeV)}{Models}} & I &  II &  III &  IV & V \\ \hline
\multirow{4}{*}{$H \to T_{4c}$} &   $0^{ +  + }$    &  $5.02 \times {10^{ - 10}}$ & $8.56 \times {10^{ - 8}}$  &  $7.72 \times {10^{ - 11}}$ & $1.60 \times {10^{ - 10}}$  & $5.08 \times {10^{ - 11}}$  \\ \cline{2-7} 
                   &   $1^{ +  - }$    & $7.33 \times {10^{ - 12}}$  & $1.18 \times {10^{ - 9}}$  & $1.03 \times {10^{ - 12}}$  & $9.16 \times {10^{ - 12}}$  &  $8.23 \times {10^{ - 13}}$ \\ \cline{2-7} 
                   &    $2^{ +  + }$   & $2.03 \times {10^{ - 11}}$  & $3.84 \times {10^{ - 9}}$  &  $3.39 \times {10^{ - 12}}$ & $7.17 \times {10^{ - 11}}$  & $3.30 \times {10^{ - 12}}$  \\ \cline{2-7} 
                   &    $tot$   & $5.29 \times {10^{ - 10}}$  & $9.07 \times {10^{ - 8}}$ & $8.16 \times {10^{ - 11}}$  & $2.41 \times {10^{ - 10}}$  &  $5.50 \times {10^{ - 11}}$ \\ \hline\hline
\multirow{4}{*}{$W^+ \to T_{4c}$} & $0^{ +  + }$      &  $1.85 \times {10^{ - 6}}$ & $3.16 \times {10^{ - 4}}$  & $2.85 \times {10^{ - 7}}$  & $5.91 \times {10^{ - 7}}$  & $1.88 \times {10^{ - 7}}$  \\ \cline{2-7} 
                   &    $1^{ +  - }$   & $2.71 \times {10^{ - 8}}$  & $4.38 \times {10^{ - 6}}$  & $3.82 \times {10^{ - 9}}$  & $3.39 \times {10^{ - 8}}$  &  $3.04 \times {10^{ - 9}}$ \\ \cline{2-7} 
                   &   $ 2^{ +  + }$   & ${\rm{7}}{\rm{.51}} \times {\rm{1}}{{\rm{0}}^{{\rm{ - 8}}}}$  &  ${\rm{1}}{\rm{.42}} \times {\rm{1}}{{\rm{0}}^{{\rm{ - 5}}}}$ & ${\rm{1}}{\rm{.25}} \times {\rm{1}}{{\rm{0}}^{{\rm{ - 8}}}}$  &  ${\rm{2}}{\rm{.65}} \times {\rm{1}}{{\rm{0}}^{{\rm{ - 7}}}}$ & ${\rm{1}}{\rm{.22}} \times {\rm{1}}{{\rm{0}}^{{\rm{ - 8}}}}$  \\ \cline{2-7} 
                   &   $tot$    &  ${\rm{1}}{\rm{.96}} \times {\rm{1}}{{\rm{0}}^{{\rm{ - 6}}}}$ & ${\rm{3}}{\rm{.35}} \times {\rm{1}}{{\rm{0}}^{{\rm{ - 4}}}}$  & ${\rm{3}}{\rm{.01}} \times {\rm{1}}{{\rm{0}}^{{\rm{ - 7}}}}$  & ${\rm{8}}{\rm{.90}} \times {\rm{1}}{{\rm{0}}^{{\rm{ - 7}}}}$  &  ${\rm{2}}{\rm{.03}} \times {\rm{1}}{{\rm{0}}^{{\rm{ - 7}}}}$ \\ \hline\hline
\multirow{4}{*}{$Z^0 \to T_{4c}$} &   $0^{ +  + }$    & ${\rm{7}}{\rm{.87}} \times {\rm{1}}{{\rm{0}}^{{\rm{ - 7}}}}$  & ${\rm{1}}{\rm{.34}} \times {\rm{1}}{{\rm{0}}^{{\rm{ - 4}}}}$  &  ${\rm{1}}{\rm{.21}} \times {\rm{1}}{{\rm{0}}^{{\rm{ - 7}}}}$ & ${\rm{2}}{\rm{.51}} \times {\rm{1}}{{\rm{0}}^{{\rm{ - 7}}}}$  &  ${\rm{7}}{\rm{.98}} \times {\rm{1}}{{\rm{0}}^{{\rm{ - 8}}}}$ \\ \cline{2-7} 
                   &    $1^{ +  - }$   & ${\rm{1}}{\rm{.15}} \times {\rm{1}}{{\rm{0}}^{{\rm{ - 8}}}}$  & ${\rm{1}}{\rm{.86}} \times {\rm{1}}{{\rm{0}}^{{\rm{ - 6}}}}$  &  ${\rm{1}}{\rm{.62}} \times {\rm{1}}{{\rm{0}}^{{\rm{ - 9}}}}$ & ${\rm{1}}{\rm{.44}} \times {\rm{1}}{{\rm{0}}^{{\rm{ - 8}}}}$  &  ${\rm{1}}{\rm{.29}} \times {\rm{1}}{{\rm{0}}^{{\rm{ - 9}}}} $\\ \cline{2-7} 
                   &    $2^{ +  + }$   & ${\rm{3}}{\rm{.19}} \times {\rm{1}}{{\rm{0}}^{{\rm{ - 8}}}}$  & ${\rm{6}}{\rm{.02}} \times {\rm{1}}{{\rm{0}}^{{\rm{ - 6}}}}$  &  ${\rm{5}}{\rm{.31}} \times {\rm{1}}{{\rm{0}}^{{\rm{ - 9}}}}$ &  ${\rm{1}}{\rm{.12}} \times {\rm{1}}{{\rm{0}}^{{\rm{ - 7}}}}$ &  ${\rm{5}}{\rm{.18}} \times {\rm{1}}{{\rm{0}}^{{\rm{ - 9}}}}$ \\ \cline{2-7} 
                   &    $tot$   & ${\rm{8}}{\rm{.31}} \times {\rm{1}}{{\rm{0}}^{{\rm{ - 7}}}}$  & ${\rm{1}}{\rm{.42}} \times {\rm{1}}{{\rm{0}}^{{\rm{ - 4}}}}$  & ${\rm{1}}{\rm{.28}} \times {\rm{1}}{{\rm{0}}^{{\rm{ - 7}}}}$  &  ${\rm{3}}{\rm{.78}} \times {\rm{1}}{{\rm{0}}^{{\rm{ - 7}}}}$ &  ${\rm{8}}{\rm{.62}} \times {\rm{1}}{{\rm{0}}^{{\rm{ - 8}}}}$ \\ \hline
\end{tabular}
\caption{Decay widths for the production of 1$S$-wave $T_{4c}$ with $J^{PC} = 0^{++},~1^{+-},~2^{++}$, and tot $(0^{++} +1^{+ -} +2^{++})$ in five distinct models through Higgs, $W^+$, and $Z^0$ decay, respectively.}
\label{tdw}
\end{table}

\begin{table}[]
\centering
\begin{tabular}{|c|c|c|c|c|}
\hline
{Models} & \multicolumn{1}{c|}{~~~~~~~$\Gamma$(GeV)} & \multicolumn{1}{c|}{Br} & \multicolumn{1}{c|}{$N_{\rm LHC}$} &$N_{\rm CEPC}$ \\ \hline
I & \multicolumn{1}{c|}{$1.96 \times {10^{ - 6}}$} & \multicolumn{1}{c|}{$9.38 \times {10^{ - 7}}$} & \multicolumn{1}{c|}{$2.89 \times {10^4}$} & $1.97 \times {10^{ 2}}$ \\ \hline
II & \multicolumn{1}{c|}{$3.35 \times {10^{ - 4}}$} & \multicolumn{1}{c|}{$1.61 \times {10^{ - 4}}$} & \multicolumn{1}{c|}{$4.93 \times {10^6}$} & $3.37 \times {10^4}$ \\ \hline
III & \multicolumn{1}{c|}{$3.01 \times {10^{ - 7}}$} & \multicolumn{1}{c|}{$1.45 \times {10^{ - 7}}$} & \multicolumn{1}{c|}{$4.44 \times {10^3}$} & $3.04 \times {10^{  1}}$ \\ \hline
IV & \multicolumn{1}{c|}{$8.90 \times {10^{ - 7}}$} & \multicolumn{1}{c|}{$4.27 \times {10^{ - 7}}$} & \multicolumn{1}{c|}{$1.31 \times {10^4}$} & $8.96 \times {10^{ 1}}$  \\ \hline
V & \multicolumn{1}{c|}{$2.03 \times {10^{ - 7}}$} & \multicolumn{1}{c|}{$9.74 \times {10^{ - 8}}$} & \multicolumn{1}{c|}{$3.00 \times {10^3}$} & $2.05 \times {10^{  1}}$ \\ \hline
\end{tabular}
\caption{Predicted decay widths ($\Gamma $), branching ratios (Br), and produced events of ${T_{4c}}$ through ${W^ + } \to {T_{4c}} + X$ at LHC and CEPC, respectively.}
\label{wdw}
\end{table}

\begin{table}[]
\centering
\begin{tabular}{|c|c|c|c|c|}
\hline
{Models} &  \multicolumn{1}{c|}{$\Gamma $(GeV)} & \multicolumn{1}{c|}{Br} & \multicolumn{1}{c|}{$N_{\rm LHC}$} & $N_{\rm CEPC}$ \\ \hline
I & \multicolumn{1}{c|}{$8.31 \times {10^{ - 7}}$} & \multicolumn{1}{c|}{$3.33 \times {10^{ - 7}}$} & \multicolumn{1}{c|}{$3.33 \times {10^{ 2}}$}& $6.82 \times {10^5}$ \\ \hline
II & \multicolumn{1}{c|}{$1.42 \times {10^{ - 4}}$} & \multicolumn{1}{c|}{$5.70 \times {10^{ - 5}}$} & \multicolumn{1}{c|}{$5.70 \times {10^4}$} & $1.17 \times {10^8}$ \\ \hline
III & \multicolumn{1}{c|}{$1.28 \times {10^{ - 7}}$} & \multicolumn{1}{c|}{$5.13 \times {10^{ - 8}}$} & \multicolumn{1}{c|}{$5.13 \times {10^{ 1}}$} & $1.05 \times {10^5}$ \\ \hline
IV & \multicolumn{1}{c|}{$3.78 \times {10^{ - 7}}$} & \multicolumn{1}{c|}{$1.51 \times {10^{ - 7}}$} & \multicolumn{1}{c|}{$1.51 \times {10^{ 2}}$} & $3.10 \times {10^5}$ \\ \hline
V & \multicolumn{1}{c|}{$8.62 \times {10^{ - 8}}$} & \multicolumn{1}{c|}{$3.46 \times {10^{ - 8}}$} & \multicolumn{1}{c|}{$3.46 \times {10^{  1}}$} & $7.09 \times {10^4}$ \\ \hline
\end{tabular}
\caption{Predicted decay widths ($\Gamma $), branching ratios (Br), and produced events of ${T_{4c}}$ through ${Z^0} \to {T_{4c}} + X$ at LHC and CEPC, respectively.}
\label{zdw}
\end{table}

\subsection{Theoretical uncertainty}

The main sources of theoretical uncertainties associated with the indirect production of $T_{4c}$ are mainly the mass of heavy quark $m_c$ and the non-perturbative effect. In Sec. \ref{basic}, we make a simple comparison of the long-distance matrix elements obtained by five distinct potential models, and analyze the large non-perturbative effects associated with the production of $T_{4c}$ tetraquarks. Next, we will analyze the theoretical uncertainty caused by the heavy quark mass $m_c=1.5 \pm 0.2$ GeV, and other input parameters are controlled at their central values.

The uncertainty of the decay widths for the indirect production of $T_{4c}$ caused by the heavy quark mass $m_c=1.5 \pm 0.2$ GeV through three decay channels is listed in Table~\ref{mcdw}. For a numerical analysis on the uncertainty, here we choose the results of long-distance matrix elements calculated by model I as representative. We can see that 
\begin{itemize}
    \item The decay widths for the production of $T_{4c}$ are $5.29\times10^{-10}$ GeV, $1.96\times10^{-6}$ GeV, and $8.31\times10^{-7}$ GeV through Higgs, $W^+$, and $Z^0$ decay, respectively.
    \item The decay widths decreases with the increase of $m_c$ due to the suppression of phase space.
     \item The theoretical uncertainty caused by $m_c=1.5 \pm 0.2$ GeV can correspondingly increase or decrease the results by almost an order of magnitude.
\end{itemize}

\begin{table}[]
\centering
\begin{tabular}{|cc|c|c|c|c|c|}
\hline
\multicolumn{2}{|c|}{\diagbox{$m_c$ (GeV)}{$\Gamma$ (GeV)}{Models}}                       & I & II & III & IV & V \\ \hline
\multicolumn{1}{|c|}{\multirow{3}{*}{$H \to T_{4c}$}} & 1.3 & $1.44 \times {10^{ -9}}$  & $2.47 \times {10^{ -7}}$  & $2.22 \times {10^{ -10}}$  & $6.56 \times {10^{ -10}}$  & $1.50 \times {10^{ -10}}$  \\ \cline{2-7} 
\multicolumn{1}{|c|}{}                  & 1.5 & $5.29 \times {10^{ -10}}$  & $9.07 \times {10^{ -8}}$  & $8.16 \times {10^{ -11}}$  & $2.41 \times {10^{ -10}}$  & $5.50 \times {10^{ -11}}$  \\ \cline{2-7} 
\multicolumn{1}{|c|}{}                    & 1.7 & $2.20 \times {10^{ -10}}$  & $3.77 \times {10^{ -8}}$  & $3.40 \times {10^{ -11}}$  & $1.00 \times {10^{ -10}}$  &  $2.29 \times {10^{ -11}}$ \\ \hline\hline
\multicolumn{1}{|c|}{\multirow{3}{*}{$W^ + \to T_{4c}$}} & 1.3 & $7.09 \times {10^{ -6}}$  & $1.21 \times {10^{ -3}}$  & $1.09 \times {10^{ -6}}$  & $3.23 \times {10^{ -6}}$  &  $7.36 \times {10^{ -7}}$ \\ \cline{2-7} 
\multicolumn{1}{|c|}{}                  & 1.5 & $1.96 \times {10^{ -6}}$  & $3.35 \times {10^{ -4}}$  & $3.01 \times {10^{ -7}}$  & $8.90 \times {10^{ -7}}$  &  $2.03 \times {10^{ -7}}$ \\ \cline{2-7} 
\multicolumn{1}{|c|}{}                  & 1.7 & $6.34 \times {10^{ -7}}$  &  $1.09 \times {10^{ -4}}$ &  $9.77 \times {10^{ -8}}$ &  $2.88 \times {10^{ -7}}$ &  $6.58 \times {10^{ -8}}$ \\ \hline\hline
\multicolumn{1}{|c|}{\multirow{3}{*}{$Z^ 0 \to T_{4c}$}}& 1.3 &  $3.01 \times {10^{ -6}}$ & $5.16 \times {10^{ -4}}$  &  $4.64 \times {10^{ -7}}$ &  $1.37 \times {10^{ -6}}$ &  $3.13 \times {10^{ -7}}$ \\ \cline{2-7} 
\multicolumn{1}{|c|}{}                  & 1.5 &  $8.31 \times {10^{ -7}}$ & $1.42 \times {10^{ -4}}$  &  $1.28 \times {10^{ -7}}$ &  $3.78 \times {10^{ -7}}$ &  $8.62 \times {10^{ -8}}$ \\ \cline{2-7} 
\multicolumn{1}{|c|}{}                  & 1.7 & $2.69 \times {10^{ -7}}$  & $4.61 \times {10^{ -5}}$  & $4.15 \times {10^{ -8}}$  &  $1.22 \times {10^{ -7}}$ &  $2.79 \times {10^{ -8                                          }}$ \\ \hline
\end{tabular}
\caption{Uncertainty of the decay widths for the production of $T_{4c}$ caused by $m_c=1.5 \pm 0.2$ GeV through three decay channels in five distinct models.}
\label{mcdw}
\end{table}

\begin{figure}
    \centering
    \includegraphics[width=0.9\linewidth]{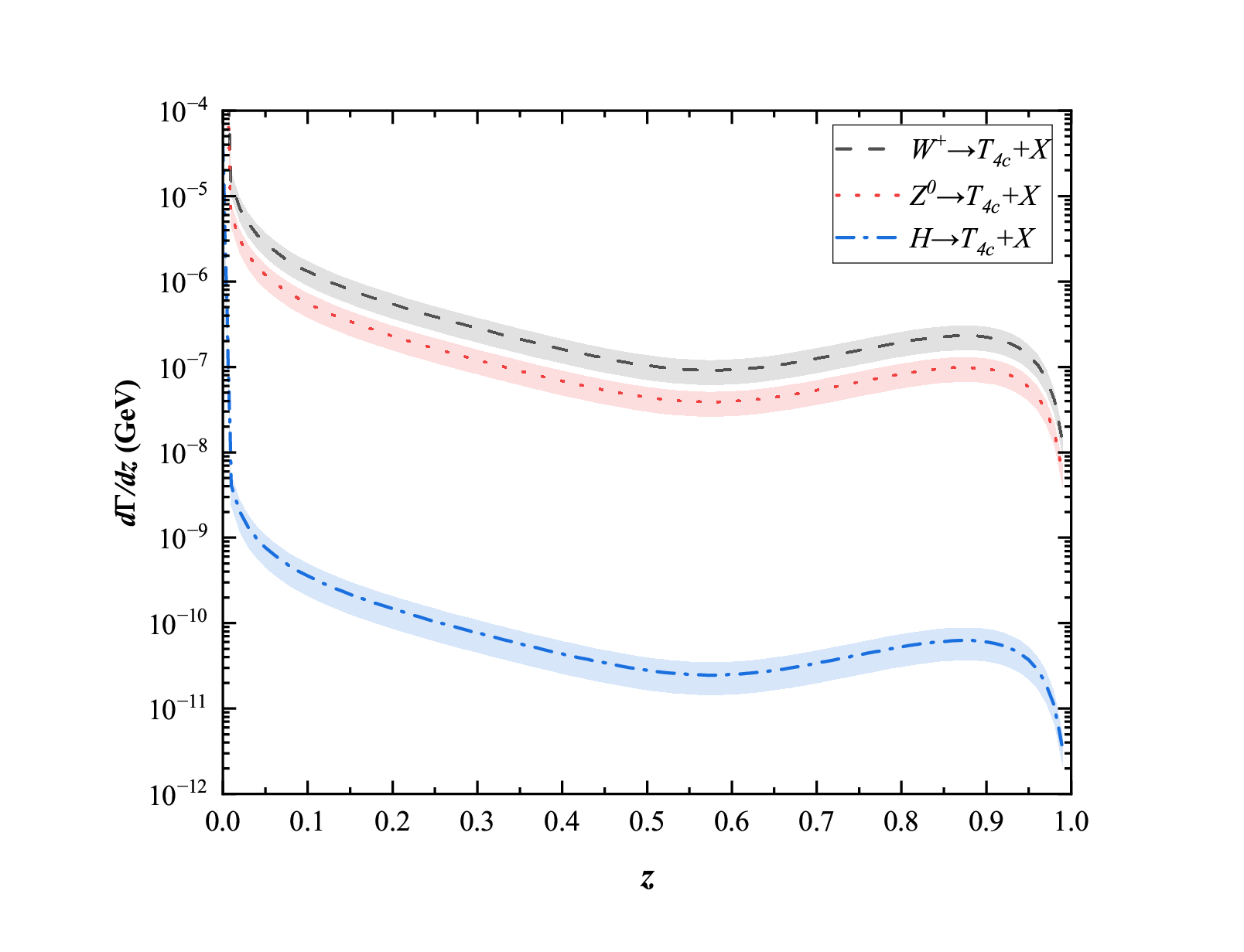}
    \caption{Energy fraction distribution of decay widths for $T_{4c}$ produced by Higgs, $W^+$, and $Z^0$ decay, respectively, with the theoretical uncertainty induced by $m_c=1.5 \pm 0.2$ GeV.}
    \label{uodw}
\end{figure}

The energy fraction $z$ distribution of decay widths for $T_{4c}$ produced by Higgs, $W^+$, and $Z^0$ decay, respectively, with the theoretical uncertainty induced by $m_c=1.5 \pm 0.2$ GeV, is plotted in Fig.~\ref{uodw}. The shape of $z$ distribution mainly depends on the fragmentation function. From Fig.~\ref{uodw}, it can be clearly seen that the $z$ distribution for the indirectly production of $T_{4c}$ through these three decay channels is similar, among which the value of decay widths by Higgs decays is significantly lower than that of $W^+$ and $Z^0$ decay, and the value through $W^+$ decay contributes the most.

\section{Summary}
The indirect production mechanism of fully charmed tetraquark $T_{4c}$ in $S$-wave is investigated mainly through Higgs$/Z^{0} \rightarrow  c +\bar {c}\rightarrow T_{4c}+ \bar {c}$ and $W^{+} \rightarrow  c +\bar{s}\rightarrow T_{4c}+ \bar{s}$ using NRQCD factorization and Suzuki approach, respectively. 
The transition probability $R$ of charm-to-$T_{4c}$ with quantum number $J^{PC}=0^{++}+1^{+-}+2^{++}$ is model-dependent, and the value of 1$S$ is 2.98$\times 10^{-7}-4.92\times 10^{-4}$, 2$S$ is 1.60$\times 10^{-7}-1.47\times 10^{-3}$. When the transverse momentum $\langle \vec{q}_T^2\rangle$ in Suzuki approaches ranges from 2.01 to 299.04 GeV$^2$, the numerical results of these two approaches are consistent with each other. 

The decay widths, branching ratios, and produced events for the production of 1$S$-wave $T_{4c}$ have been predicted at LHC and CEPC, respectively. The results in five distinct models can differ by 3-4 orders of magnitude, with the smallest result coming from model V and the largest from model II. The largest contribution of branching ratios for $T_{4c}$ production is via $W^+$ decay, which is $9.74\times 10^{-8}\sim 1.61\times 10^{-4}$. Given that at LHC, the produced number of $T_{4c}$ events through $W^+$ decay is $3.00\times10^3 \sim 4.93\times 10^6$, there is some potential to observe it at LHC. Due to the abundance of $Z^0$-events produced at CEPC, a sufficient number of $T_{4c}$ events are produced through $Z^0$ decays, it can be $10^{4-8}$ with different potential models, which is likely to be detected in future experiments. However, the decay widths and predicted produced events of $T_{4c}$ by Higgs decay is very small, even negligible, compared to that through $W^+$ and $Z^0$ decay. 

The theoretical uncertainty caused by the heavy quark mass $m_c$ is analyzed, including the decay widths and the energy fraction distribution of decay widths. For a numerical analysis on the uncertainty, here we choose the results of long-distance matrix elements calculated by model I as representative. The decay widths for the production of $T_{4c}$ decreases with the increase of $m_c$ due to the suppression of phase space. The theoretical uncertainty caused by $m_c=1.5 \pm 0.2$ GeV can correspondingly increase or decrease the results by almost an order of magnitude. The $z$ distributions for the indirectly production of $T_{4c}$ through these three decay channels are similar, among which the value of decay widths by Higgs decays is significantly lower than that of $W^+$ and $Z^0$ decay, and the value through $W^+$ decay contributes the most.

{\bf Acknowledgments:} This work was partially supported by the Natural Science Foundation of Guangxi (no. 2024GXNSFBA010368 and no. 2025GXNSFAA069775). This work was also supported by the Central Government Guidance Funds for Local Scientific and Technological
Development, China (no. Guike ZY22096024) and the National Natural Science Foundation of China (no. 12005045).

\end{document}